\documentstyle[preprint,aps,epsf,amssymb]{revtex}

\tighten
\draft

\def\ol{\overline}

\def\divv{\text{div}\, }

\def\gmu{\gamma^\mu}

\newcommand{\dpartial}[2]{\frac{\partial {#1}}{\partial {#2}}}

\newcommand{\x}[1]{x_{{#1}}}
\newcommand{\X}[1]{X_{{#1}}}
\renewcommand{\j}[1]{j_{{#1}}}

\newcommand{\bq}[1]{{\bf q}_{#1}}
\newcommand{\bQ}[1]{{\bf Q}_{#1}}

\def\a{\alpha}
\def\g{\gamma}
\renewcommand{\rho}{\varrho}
\def\C{{\Bbb C}}
\def\R{{\Bbb R}}
\def\F{{\cal F}}

\def\Li{Lorentz invariant}
\def\LI{Lorentz invariance}
\def\BM{Bohmian mechanics}

\def\qt{quantum theory}

\def\rel{relativistic}

\begin{document}

\title{Hypersurface Bohm-Dirac models}

\author{Detlef  D\"{u}rr}
\address{Mathematisches Institut der Universit\"{a}t
M\"{u}nchen,\\ Theresienstra{\ss}e 39, 80333 M\"{u}nchen,
Germany}
	\author{Sheldon Goldstein}
\address{Department of Mathematics,
	Rutgers University,\\ New Brunswick, NJ 08903, USA}
\author{Karin M\"{u}nch-Berndl}
\address{Institut f\"ur Angewandte Mathematik, Universit\"at
Z\"urich-Irchel,\\ Winterthurer Strasse 190,
         8057 Z\"urich, Switzerland}
\author{Nino Zangh\`{\i}}
\address{Dipartimento di Fisica, Universit\`a di Genova,
Sezione INFN Genova,\\ Via Dodecaneso 33, 16146 Genova, Italy}
\date{May 28, 1999}
\maketitle

\begin{abstract} We define a class of \Li\ Bohmian quantum
models for $N$ entangled but noninteracting Dirac particles.
Lorentz invariance is achieved for these models through the
incorporation of an additional dynamical space-time structure
provided by a foliation of space-time. These models can be
regarded as the extension of Bohm's model for $N$ Dirac
particles, corresponding to the foliation into the equal-time
hyperplanes for a distinguished Lorentz frame, to more general
foliations. As with Bohm's model, there exists for these
models an equivariant measure on the leaves of the foliation.
This makes possible a simple statistical analysis of position
correlations analogous to the equilibrium analysis for (the
nonrelativistic) Bohmian mechanics.
\end{abstract}

\pacs{03.65.Bz}

\narrowtext

\section{Introduction}

Among the different approaches to resolving the conceptual
problems of
\qt, Bohm's approach is perhaps the simplest.  In a nutshell,
it consists in adding the most basic dynamical variables,
obeying additional evolution equations, to the description of
a quantum system provided by its wave function $\psi$.  For
nonrelativistic \qt\ the additional variables are the
positions of the particles, which evolve according to a
``guiding equation'' naturally suggested by the Schr\"odinger
evolution.  This theory---usually called \BM{} or the
pilot-wave theory---is well understood. It has been analyzed,
and its connection with the predictions of orthodox quantum
theory explained, in the original papers of Bohm
\cite{Bohm52} as well as in later works (see, e.g.,
\cite{Bell,DGZ92a,CFG}).  One of the main problems remaining
for the Bohmian (or any other) approach is to find a {\it
satisfactory \/} \rel\ quantum theory, a theory that is fully
Lorentz invariant while avoiding the profound conceptual
difficulties of orthodox quantum theory.

In his original papers, Bohm had an outline for a ``Bohmian''
field theory, with fields on space-time as the additional
variables. A year later he proposed a ``Bohmian'' model for
one Dirac particle
\cite{Bohm53}, which was subsequently extended by Bohm and
coworkers to
$N$ Dirac particles
\cite{BohmHiley}.  For this $N$-particle model the additional
variables are, as in Bohmian mechanics, the positions ${\bf
Q}_k,\ k=1,\dots,N$, of the particles. However, in contrast
with Bohmian mechanics, the guiding equation for this theory

\begin{equation}
\frac {d{\bf Q}_{k}}{dt} = \frac {\psi ^\dagger \bbox{\a}_{k}
\psi}{\psi ^\dagger \psi}   \label{BohmDirac}
\end{equation} is ultralocal on configuration space: The right
hand side of (\ref{BohmDirac}) depends only upon the value of
$\psi$ at the positions of the particles and not upon spatial
derivatives of $\psi$ there. Here $\psi= \psi(\bq{1}, \dots ,
\bq{N}, t)$, taking values in the
$N$-particle spin space $(\C^4)^{\otimes N}$, solves the
$N$-particle Dirac equation ($\hbar = c =1$)
\begin{eqnarray} i \dpartial{\psi}{t}  =   \sum_{k=1}^N
\Bigl( -i\bbox{\a}_{k} \cdot
\bbox{\nabla}_{k}  - e  \bbox{\a}_{k} \cdot {\bf A}(\bq{k},t)
 & & \nonumber\\ + e \Phi(\bq{k},t)  +\beta_{k} m  \Bigr) \psi
,& & 
\label{DiracN}
\end{eqnarray} where
$\bbox{\a}_{k}=(\a_{k}^1,\a_{k}^2,\a_{k}^3)$,
$\a_{k}^i = I\otimes \cdots \otimes I\otimes \a^i \otimes
I\otimes
\cdots \otimes I$, with the $i$-th Dirac $\alpha$ matrix
$\a^i$ at the
$k$-th of the $N$ places,
 and $\beta_{k}$ is defined analogously.  $\Phi$  and $ {\bf
A}$  are external electromagnetic potentials.  (We may of
course consider particle-dependent masses $m_{k}$, charges
$e_{k}$, and external potentials $\Phi_{k}$ and $ {\bf
A}_{k}$, but for simplicity we shall not do so.)  We shall
call this model the  Bohm-Dirac model (BD model).  Just as
with Bohm's proposal for a  field theory, the BD model
requires for its  formulation the specification of a
distinguished frame of reference---in terms of which the
actual configuration  $(\bQ{1},
\dots , \bQ{N})$ and the generic configuration $(\bq{1}, \dots
, \bq{N})$ at time $t$ is defined---and in fact the model is
not \Li\ if $N>1$
\cite{BohmHiley}.

However, for $N=1$  this model is \Li,  and may be formulated
in a covariant way: Writing $X=X(\tau)$ for the space-time
point along a trajectory, with (scalar) parametrization
$\tau$, the guiding equation may be written  as
\begin{equation}
\label{BD1particle}
\frac {dX}{d\tau} = j \equiv  \ol \psi \gamma \psi
\end{equation}  with $\psi$ satisfying the Dirac equation
\begin{equation}
	 \left( i \gamma\cdot \partial - e \gamma\cdot A -m
	\right) \psi =0 ,
 	\label{eq:onepDirac}
 \end{equation}  where $\gamma\cdot \partial\equiv\gmu
\partial_{\mu}$ and $\gamma\cdot  A\equiv \gamma^\mu
A_\mu(x)$. Note that the right hand side of
(\ref{BD1particle}), the Dirac current $j=j^{\mu}\equiv\ol \psi
\gamma^{\mu} \psi$, is the simplest 4-vector that can be
constructed {}from the Dirac spinor $\psi$.

Note also that the parameter $\tau$ has no intrinsic physical
significance, so that equation (\ref{BD1particle}) is
equivalent to
$$
\frac {dX}{d\tau} = aj
$$ with arbitrary positive scalar field $a=a(x)$.  It is not
the field of
 4-vectors $j$ (having direction and length) that determines
the particle
 motion, but rather the field of directions defined by $j$. In
other words,
 the law for the particle motion could be formulated in a
purely
 geometrical manner as the condition that the Dirac current
$j$ at every
 point along the trajectory be tangent to the trajectory at
that point.

Because the Dirac current is time-like and divergence
free,\footnote{The claims in this and the next paragraph
follow directly {}from the application of the divergence
theorem (or Stokes' theorem) to an infinitesimally thin tube
of paths between $\Sigma_0$ (see below) and the relevant
hypersurface
$\Sigma$.}
$$
\partial\cdot j = 0,
$$ there is a dynamically distinguished probability
distribution on the set of particle paths $X(\tau)$ arising
{}from (\ref{BD1particle}).  Any  distribution on this space
of  paths can be defined by specifying for the path the
crossing probability for some given equal-time surface
$\Sigma_0$ in some Lorentz frame. (By this crossing
probability we mean the distribution of the point through
which the path crosses $\Sigma_0$, which is the same thing as
the probability distribution for the position of the particle
in this frame at the given time.)  The distinguished
distribution is then defined by the crossing probability for
$\Sigma_0$ given by $\rho = j^0 = \psi ^\dagger \psi$ on
$\Sigma_0$ (with $\psi$ suitably normalized), which can be
written in a covariant manner as $j\cdot n$ where $n$ in the
future-oriented unit normal to the surface.  For this
distribution the crossing probability for any other equal-time
surface will also be given by
$j\cdot n$, both for the original frame and any other Lorentz
frame. We may roughly summarize the situation by saying that
for the distinguished probability distribution, quantum
equilibrium holds  in all Lorentz frames at all times, with
the quantum equilibrium distribution given by
$\rho = \psi ^\dagger \psi$.

More generally, the crossing probability for any space-like
hypersurface
$\Sigma$ will also be given by $j\cdot n$, with $n=n(x)$ the
future-oriented unit normal field to $\Sigma$.  Moreover, for
any oriented hypersurface
$\Sigma$, the crossing measure (a signed measure that need not
be normalized), which describes the expected number of signed
crossings through area elements of $\Sigma$, with negatively
oriented crossings counted negatively, is, for the
distinguished distribution, also given by
$j\cdot n$, with $n=n(x)$ now the positively oriented unit
normal field to
$\Sigma$.\footnote{In this regard it is perhaps worth noting
the following: In Minkowski space there is a natural duality
between divergence-free vector fields and closed 3-forms. Such
a vector field defines a ``deterministic'' random  path, whose
``law'' is given directly by the vector field, as in
(\ref{BD1particle}), and whose statistics are governed by the
dual 3-form, in the manner just described.}

The $N$-particle BD model (\ref{BohmDirac}) also has a 
dynamically distinguished probability distribution on paths. 
As a consequence of (\ref{DiracN}) $\rho = \psi ^\dagger
\psi$ satisfies, in the Lorentz frame in which the dynamics is
defined, the continuity equation
\begin{equation}
\label{conteqnBD}
\dpartial{\rho}{t} + \sum_{k=1}^N \bbox{\nabla} _{k} \cdot
{\bf J}_{k} = 0,
\end{equation} where
\begin{equation}
	{\bf J}_{k}= \rho {\bf v}_{k} = \psi ^\dagger \bbox{\a}_{k}
	\psi\,.
	\label{eq:bdj}
\end{equation}
 Thus, if the joint probability distribution for the positions
of the $N$ particles is given $\rho = \psi ^\dagger \psi$ at 
some time $t= t_0$, then, for the corresponding distribution
on paths, it will be given by
$\rho = \psi ^\dagger \psi$ at all times $t$. However, even
for this distinguished distribution, quantum equilibrium will
not in general hold in other Lorentz frames: The joint
distribution of crossings of equal-time surfaces for other
frames will in general not be given by
${\psi'}^\dagger \psi'$ (where $\psi'$ is the wave function in
the relevant Lorentz frame)\cite{Samols,eprli}. Nonetheless,
Bohm and coworkers have argued that the observational content
of this model is as
\Li\ as the covariant formalism of \rel\ \qt: Since the
predictions for results of measurements for this model can be
regarded as reflected in the configuration of various devices
and registers---and hence can be derived  {}from probabilities
for positions given by $\rho = \psi ^\dagger
\psi$---at a common time in the distinguished frame, these
predictions must agree with those of the usual interpretation.
Thus no violation of
\LI\  can be detected in experiments \cite{BohmHiley}. (In
particular, the identity of the distinguished Lorentz frame
cannot be ascertained by means of any possible observation.)

Lorentz invariance is, however, a delicate issue.  Indeed, any
theory can be made trivially Lorentz invariant (or invariant
under any other space-time symmetry), even on the microscopic
level, by the incorporation of suitable additional structure
\cite{eprli}.  For this reason  Bell has stressed that one
should consider what he has called ``serious Lorentz
invariance,'' a notion, however, that is extremely difficult
to make precise in an adequate way \cite{Bell}.  Lacking a
general criterion, we may  nonetheless begin to  get a handle
on ``serious  Lorentz invariance'' by analyzing some specific 
models. If the models involve additional structure, then
whether or not we have serious Lorentz invariance will depend,
of course, upon the detailed nature of this structure.

In \cite{eprli} we have considered a model for which the
additional structure for a system of $N$ (noninteracting)
Dirac particles is provided by a global synchronization among
the particles: The trajectories of the particles are such that
each one of them at some given space-time point is tangent to
a vector field determined, given the wave function, by that
point and those  points along the  trajectories of the other
particles with which that point has been ``synchronized.''
This additional synchronization  structure is defined
implicitly by the equation of motion and the model  is not
amenable to a statistical analysis in any obvious way.  In 
other words, this model is not statistically transparent (see
Section  IV of \cite{eprli}).  Nonetheless, even this model 
provides  a counterexample to the widely held belief that a
Lorentz invariant Bohmian theory  for many particles is
impossible (unless only  product states are allowed).  In this
regard, see also the local model of  Squires~\cite{sqmodel}.

In this paper we shall analyze a statistically transparent
counterexample, the ``hypersurface Bohm-Dirac model'' (HBD
model).   The basic idea was proposed in \cite{DGZ90} in the
context of  bosonic quantum field theory: In addition to the
wave function and field variables, a  distinguished foliation
of space-time---a new element of geometrical structure 
defining simultaneity surfaces---is suggested as an additional
dynamical variable  of the theory.  These surfaces need not be
hyperplanes.  The defining  (\Li ) equations of the theory
should describe the evolution of  the wave function, the 
field variables, and the simultaneity surfaces.  For a careful
philosophical  discussion of how this may be compatible with
some appropriate notion of  relativity, even if the
simultaneity surfaces should turn out to be  unobservable, see
Maudlin \cite{Maudlin}.

Here we shall consider such a theory, not for fields but for
$N$ (noninteracting) Dirac particles.  We shall discuss an as
yet incomplete hypersurface Bohm-Dirac model: The law for the
evolution of the foliation is not specified, beyond the
requirement that it not involve the positions of the
particles.  We present no hypothesis concerning the origin of
the foliation, but have in mind that the foliation should
ultimately be governed by a Lorentz invariant law, one that
may, for example, involve the
$N$-particle wave function.  (For definiteness we shall give
some very tentative and less than compelling examples of laws
for the foliation in Section \ref{sect:persp}.)  However, we
show in Subsection
\ref{subs:qe} that, regardless of how the foliation is
determined, the dynamics of the HBD model preserves the
quantum equilibrium distribution on the leaves of the
foliation.  Thus the model is amenable to the same sort of
statistical analysis as for nonrelativistic Bohmian mechanics.
This is discussed briefly in Subsection \ref{subs:qm}.

\section{The hypersurface Bohm-Dirac model}\label{sect:surf}

A general foliation $\F$ of codimension one on Minkowski space
$M$ can approximately be thought of as a partition of $M$ into
3-dimensional
 hypersurfaces. These hypersurfaces are the leaves of the
foliation. The simplest way to obtain a foliation is by a
smooth function
$f:M\to \R$ without critical points, i.e., $df\neq 0$
everywhere. The level sets $f^{-1}(s)$ are smooth
hypersurfaces and form a foliation of
$M$. With the one-form $df_x$, which vanishes on the tangent
space of the hypersurface through $x\in M$, we may associate
by the Lorentz metric the normal vector field
$\partial f(x)$. If this is time-like everywhere, and thus the
foliation hypersurfaces space-like, we may normalize
$\partial f(x)$ to obtain a unit normal vector field
$n(x)$ associated with the foliation $\F$.

We shall consider in this paper only space-like foliations,
i.e., foliations by space-like hypersurfaces. While obviously
different $f$'s may generate the same foliation $\F$, the
future-oriented unit normal vector field
$n$ is uniquely determined by $\F$. When does a vector field
$v(x)$ determine a foliation $\F$ such that for all $x\in M$,
$v(x)$ is normal to the tangent space of the foliation
hypersurface through $x$? If we denote by
$V$ the one-form associated with $v$ by the Lorentz metric,
then, by Frobenius' theorem, the necessary and sufficient
condition is that $V$ be completely integrable, $V\wedge dV
=0$.

Apart {}from the foliation, the other dynamical variables of
the hypersurface Bohm-Dirac model are the usual ones: the wave
function
$\psi$, here for $N$ Dirac particles, and the $N$-path, the
$N$-tuple of (everywhere either time-like or light-like)
space-time paths, which describes the trajectories of the $N$
Dirac particles.  Covariant laws for these dynamical variables
suggest themselves when we write those of the the Bohm-Dirac
model, defined by (\ref{BohmDirac}) and (\ref{DiracN}), in a
coordinate-free, i.e., covariant manner.

To achieve this we consider first of all the $\psi$-function
in the multi-time formalism: For $N$ Dirac particles the wave
function $\psi=
\psi (x_{1}, x_{2}, \dots , x_{N})$, $x_{k}\in M$, takes
values in  the
$N$-particle spin space $(\C^4)^{\otimes N}$ and satisfies
$N$  Dirac equations
\begin{equation}
\label{Diraceqn}
\left( i \gamma_{k}\cdot \partial_{k} - e \gamma_{k}\cdot 
A(x_{k}) -m
\right) \psi =0 \,,
\end{equation} $k=1,\dots , N$. Here $\gamma_{k} = I\otimes
\cdots \otimes I\otimes \gamma
\otimes  I\otimes \cdots \otimes I$, with $\gamma$ at the
$k$-th of the
$N$  places, and $A$ is an external electromagnetic
potential.  (Just as with (\ref{DiracN}), we may of course
consider particle-dependent  masses
$m_{k}$, charges $e_{k}$, and external potentials $A_{k}$.) 
The  system of equations (\ref{Diraceqn}) is a covariant
version of (\ref{DiracN}); in this multi-time form the \LI\ of
the law for $\psi$ is  manifest
\cite{BohmHiley}.\footnote{Note that in the single-time form
(\ref{DiracN}) we can easily add an explicit interaction
potential
$V(\bq{1},\dots , \bq{N},t)$ for the $N$ Dirac particles,
while in the multi-time form this is impossible.} The $N$
Dirac particles are  coupled by the common wave function
$\psi$.  If this is entangled, we  have nonlocal correlations
between the $N$ particles, despite the fact that the particles
are noninteracting.

We shall now develop the guiding law for the $N$-path. Note
that the numerator of the right hand side of
(\ref{BohmDirac})   is given by a current $j_k$,
$$
 j_k = \overline{\psi}
\gamma_1^0\ldots\gamma_k\ldots
\gamma_N^0\psi ,
$$ that involves matrix elements of an  operator having as
factors the 0-component $\gamma^0$ of a 4-vector for all but
the $k$-th particle. Therefore $j_k$ can be expressed in a
covariant manner by replacing
$\gamma_k^0$ in the above expression with $\gamma_k\cdot n$,
where $n$ is the future-oriented unit normal to the $t$ =
const hyperplanes,
\begin{equation}
	j_k = \overline{\psi} (\gamma_1\cdot n)\ldots \gamma_k \ldots
	(\gamma_N\cdot n)\psi .
	\label{eq:flaths}
\end{equation}  Moreover, the denominator of the right hand
side of (\ref{BohmDirac}) can be  expressed covariantly as
$j_k\cdot n$.  Then the covariant velocity of  the $k$-th
particle---with respect to the time of a Lorentz frame with
$n$ as time axis---is
\begin{equation} \frac{dX_k}{dt}= \frac{j_k}{j_k\cdot n}
.\label{eq:BDpara}
\end{equation}   Since $j_k\cdot n=\overline{\psi}
(\gamma_1\cdot n)\ldots (\gamma_N\cdot n)\psi $ is independent
of $k$, we may reparametrize the paths with a parameter $s$ so
related to $t$ that $t'(s) = j_k\cdot n$ to obtain
\begin{equation}
\frac{dX_k}{ds}= j_k .
\label{eq:s}
\end{equation} More generally, by  further reparametrization,
we may obtain ${dX_k}/{d\tau}= a j_k $,  where
$a$ is any positive scalar field.  The physical particle
dynamics---i.e., the $N$ space-time paths defined by the
equations of motion (and initial conditions)---is  invariant
under reparametrization.

A manifestly ``parametrization invariant'' formulation of the
dynamics---that is, such that a time parameter  plays no
role---is easily obtained: The space-time paths for the $N$
particles are constrained by the currents $j_k$ by requiring
that the path for the
$k$-th  particle at the point $x_k$ be tangent to the current
$j_{k}$ evaluated at $x_{k}$ and at the the intersection
points of the
 paths of the $N-1$ other particles with the $t=$
const-hyperplane
$\Sigma_t$ containing
$x_{k}$. If we denote by $X_k(\Sigma_t)$ the intersection
point of the path $X_k$ with the hyperplane $\Sigma_t$, and by
$\dot X_k(\Sigma_t)$ a tangent of (or the tangent line to) the
path $X_k$ at $X_k(\Sigma_t)$, we may write the law for the
$N$-path as
\begin{equation}\label{eq:lawBD}
\dot X_k(\Sigma_t) \parallel j_k\left( X_1(\Sigma_t), \dots
,X_N(\Sigma_t)\right)  ,
\end{equation} using the symbol $\parallel$ for ``is parallel
to.'' In this geometric formulation the Bohm-Dirac dynamics
depends upon the Lorentz frame only via its associated
foliation into simultaneity hypersurfaces $\Sigma_t$, and thus
naturally extends to an arbitrary foliation $\F$ of Minkowski
space-time $M$ by curved space-like
hypersurfaces:\footnote{This is in marked contrast with the
parametrized dynamics such as given by equations
(\ref{eq:BDpara}) or (\ref{eq:s}), which need not extend in
anything like the same form to a general foliation since the
parametrized paths generated by the dynamics need not, in
general, respect the foliation.}  

\begin{figure}
\begin{center}
\epsfxsize=8.5cm
\epsfbox{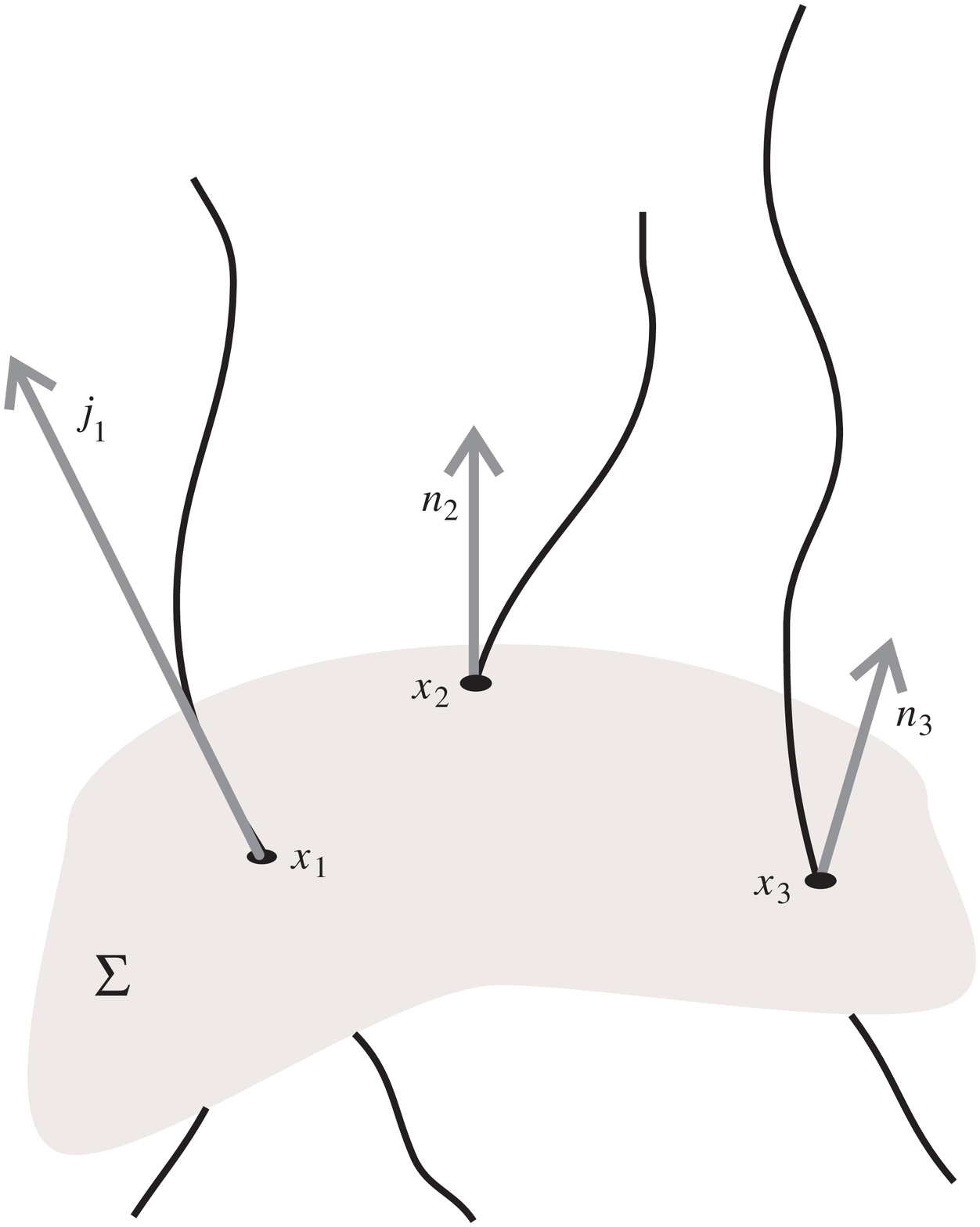}
\end{center}
\caption{Geometrical formulation of the dynamics for a system
of three particles: For each particle the path of that
particle, say particle 1 at
$x_{1}$, must be tangent to the 4-vector $j_{1}$ which is
determined by: 1) the intersections $x_{2}$ and $x_{3}$ of the
trajectories of the other two particles with the hypersurface
$\Sigma$ containing $x_{1}$, 2) the future-oriented unit
normals $n_{2}$ and $n_{3}$ at these points, and 3) the wave
function of the system evaluated at $x_{1}$, $x_{2}$ and
$x_{3}$:
$ j_1 = \overline{\psi}(x_1,x_2,x_3) \gamma_1 (\gamma_2 \cdot
n_2) (\gamma_3\cdot n_3)\psi (x_1,x_2,x_3).$}
\label{fig:path}
\end{figure}

Given such a foliation $\F$ and $\Sigma \in \F$, let
$X_k(\Sigma)$ be the intersection of the path $X_k$ with
$\Sigma$,\footnote{Note that the paths
$X_k$ comprising an $N$-path, since they are nowhere
space-like, can intersect $\Sigma$ at most once. This is the
main reason why it is important that the foliation $\F$ be
space-like. Of course, also {}from the physical point of view
a synchronization along space-like hypersurfaces yields a
picture which perhaps makes most sense. We shall assume,
without further ado, global existence: that a fragment of an
$N$-path locally satisfying the HBD tangency condition, see
(\ref{eq:lawHBD}), can be continued in such a manner that each
of its paths $X_k$ intersects every
$\Sigma \in \F$.} and let $\dot X_k(\Sigma)$ be a tangent of
(or the tangent line to) the path $X_k$ at $X_k(\Sigma)$. The
law of the $N$-path
$X=(X_1, \dots, X_N)$ for the hypersurface Bohm-Dirac model is
defined by the currents $j_k$ naturally extending
(\ref{eq:flaths})
\begin{equation}
	 j_k = \overline{\psi} (\gamma_1\cdot n_{1})\ldots \gamma_k
\ldots
	 (\gamma_N\cdot n_N)\psi ,
\label{eq:flhs}
\end{equation} where $n_{1}\equiv n(x_{1}), \ldots, n_{N}\equiv
n(x_{N})$,  with $n$ the future-oriented unit normal vector
field associated with $\F$, via the HBD tangency condition 
(see also Fig.\
\ref{fig:path})
\begin{equation}\label{eq:lawHBD}
\dot X_k(\Sigma) \parallel j_k\left( X_1(\Sigma), \dots
,X_N(\Sigma)\right)  .
\end{equation} [By considering the action of a suitable Lorentz
transformation on $\g^0\g
\cdot n$ for arbitrary time-like unit vector $n$ (transforming
$n$ to
$(1,0,0,0)$), one sees that $\g^0\g \cdot n$ is a positive
operator in spin space $\C^4$. Hence $(\g^0_1\gamma_1\cdot
n_{1}) \ldots (\g^0_k\gamma_k\cdot n) \ldots
(\g^0_N\gamma_N\cdot n_N)$ is also positive, i.e., $j_k\cdot
n\geq 0$ with ``='' only if $\psi=0$. This means that, where
it is nonzero, $j_k$ is future-oriented and, like the path
$X_k$, nowhere space-like.]

We may also write down the equations of motion in the
parametrized form analogous to (\ref{eq:BDpara}) or
(\ref{eq:s}). To do so it is convenient to label the
hypersurfaces of the foliation using a function $f:M\to \R$
that generates the foliation as described above, and use this
hypersurface labeling as the parameter for the particle
trajectories---so that
$\X{k}(s)$ is on the hypersurface $f^{-1}(s)$. {}From the
geometrical characterization of the dynamics (\ref{eq:lawHBD})
we know that
${dX_k}/{ds}$ is parallel to $j_k(X_1(s), \dots ,X_N(s))$, and
the scale factor required to ensure $f(\X{k}(s))=s$ for all
$k$ and $s$ is easily seen to be $1/(\partial f \cdot j_{k})$.
Therefore
\begin{equation}
\label{equ:dyn}
\frac{dX_k}{ds} =\frac{j_{k}(X_1(s), \dots ,X_N(s))}{\partial f
(X_k(s))\cdot j_{k} (X_1(s), \dots ,X_N(s))} .
\end{equation} For a flat foliation we may choose  a Lorentz
frame such that the foliation hyperplanes are the $x^0=$
const-planes, i.e.
$f(x) =x^0$ for all
$x$. Then $n= \partial f  = (1,0,0,0)$ and (\ref{equ:dyn})
reduces to the Bohm-Dirac law (\ref{BohmDirac}).

\section{Statistical analysis of the HBD
model}\label{sect:statan}

\subsection{Quantum equilibrium}\label{subs:qe}

We shall show now that for the hypersurface Bohm-Dirac model,
with foliation $\F$, there is a distinguished probability
measure on $N$-paths
$X$ satisfying the HBD tangency condition (\ref{eq:lawHBD}),
one for which the distribution of hypersurface crossings
$X_1(\Sigma), \dots ,X_N(\Sigma)$ for $\Sigma\in\F$ depends
only upon $\psi$ restricted to
$\Sigma$ (or, more precisely, to $\Sigma^N$) for $\psi$
satisfying (\ref{Diraceqn}). We shall say that such a
distinguished measure, as well as the corresponding
hypersurface crossing distribution, is {\it equivariant,\/}
defining {\it quantum equilibrium\/}.  The physical
significance of the hypersurfaces $\Sigma\in\F$ is thus
twofold: They serve (via (\ref{eq:lawHBD})) to define the
motion of the particles, and, for a quantum equilibrium
$N$-path, it is ``on these hypersurfaces'' that, manifestly,
the ``particles are in quantum equilibrium.''

The natural candidate for the equivariant crossing probability
density
$\rho$ of the HBD model is given by the obvious covariant
extension of the equivariant density $ \psi^{\dagger}\psi \ (=
\overline{\psi}
\gamma_1^{0}\ldots \gamma_N^{0} \psi)$ of the BD model:
\begin{equation}
	\rho = \overline{\psi} (\gamma_1\cdot n_{1})\ldots
(\gamma_N\cdot
	 n_N)\psi .
	\label{eq:covprop}
\end{equation}   To see that this is in fact equivariant, note
the following: In view of (\ref{eq:flhs}), (i) $\rho =
j_{k}\cdot n_{k}$ and
\begin{equation}
	j_{k}\cdot n_{k}\text{ is independent of }k.
	\label{eq:bas}
\end{equation}
 Furthermore, (ii) the currents $\j{k}$ are divergence free:
\begin{equation}
\partial_k\cdot \j{k} = 0,
\label{eq:divfree}
\end{equation} which follows immediately {}from
(\ref{eq:flhs})  using the Dirac equation (\ref{Diraceqn}) and
its adjoint. These two properties of the currents,
(\ref{eq:bas}) and  (\ref{eq:divfree}), are the key
ingredients for the proof of the equivariance of $\rho$. For
any current satisfying (\ref{eq:bas}) and 
(\ref{eq:divfree}),  for the particle  dynamics defined by
(\ref{eq:lawHBD}), $\rho = j_{k}\cdot n_{k}$ is an equivariant
probability density for crossings of the leaves of the
foliation.\footnote{In contrast, the current $j_k=
\ol\psi\g_k\psi$ we considered in \cite{eprli} satisfies
(\ref{eq:divfree}) but not (\ref{eq:bas}).}

The proof of this assertion consists of two steps: First we
determine how an arbitrary probability density $R$ on
crossings of a foliation hypersurface $\Sigma$ evolves under
the dynamics (\ref{eq:lawHBD}), i.e., we formulate the
continuity equation of the hypersurface dynamics. In the
second step, we show that $R=\rho$ solves the continuity
equation. It then follows that if the probability distribution
of the ``positions of the $N$ particles'' on $\Sigma\in\F$ is
given by $\rho$ restricted to $\Sigma$, then for any other
hypersurface $\Sigma'\in\F$, the probability distribution of
the ``positions of the $N$ particles'' on $\Sigma'$ which
emerges by transport according to the dynamics
(\ref{eq:lawHBD}) is given by $\rho$ restricted to $\Sigma'$.
Thus $\rho$ is equivariant.

\begin{figure}
\begin{center}
\epsfxsize=8.5cm
\epsfbox{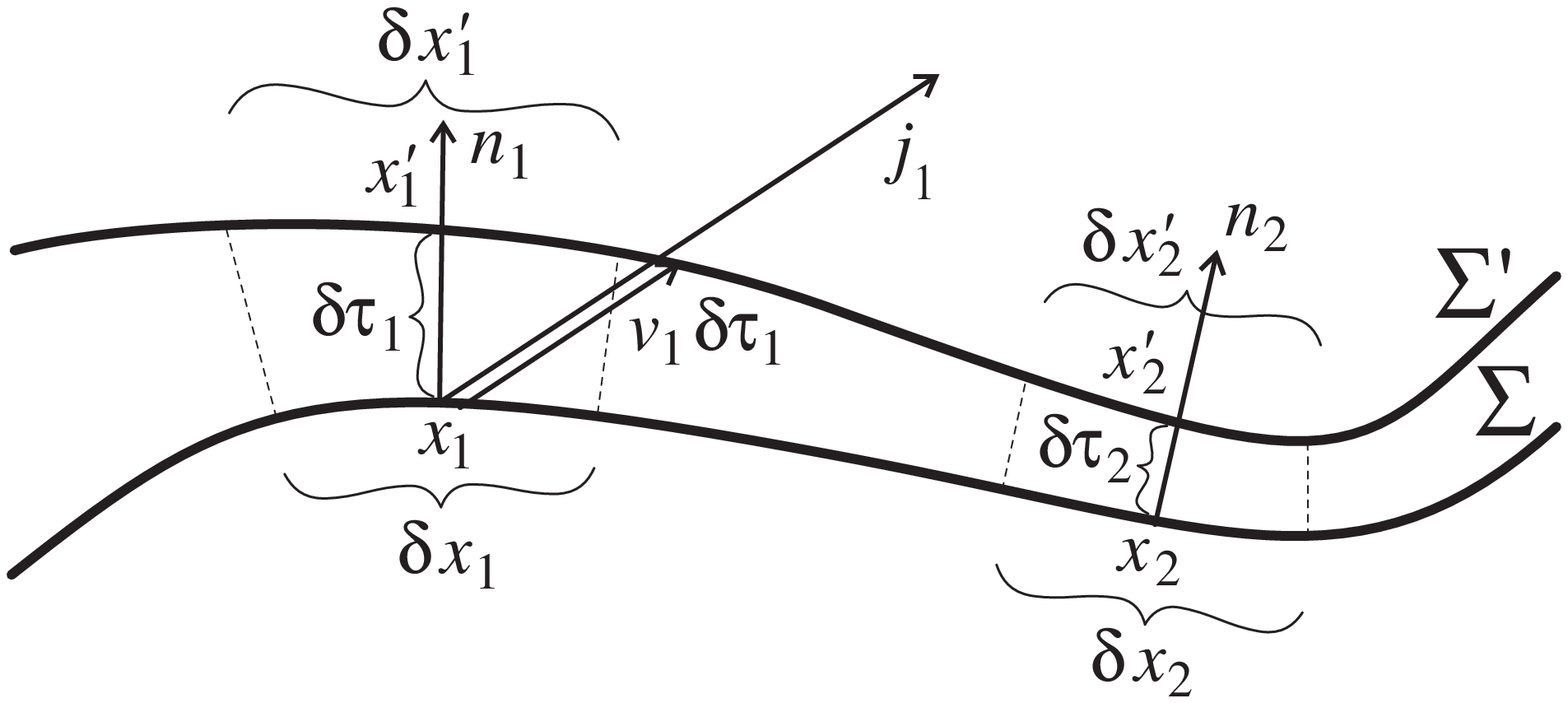}
\end{center}
\caption{Motion of two particles in  one space dimension {}from
hypersurface $\Sigma$ to $\Sigma'$:  space-time view. We have
indicated the positions of the primed points $x_k'$ obtained
{}from $x_k$ via displacement {}from $\Sigma$ to $\Sigma'$ in
the normal direction, and the images $\delta x_k'$ of the
regions
$\delta x_k$ under this correspondence. The point on $\Sigma'$
to which particle $k$ moves when starting at $x_k\in \Sigma$
is given (to leading order) by $x_k + v_k
\delta \tau_k$ with $v_k = j_k/ (j_k\cdot n_k)$, where 
$\delta \tau_k$ is the Minkowski distance between $x_k$ and
$x_k'$.}
\label{fig:st}
\end{figure}

Consider thus two infinitesimally close hypersurfaces $\Sigma$
and
$\Sigma'$ belonging to the foliation $\F$. The probability
distribution of the positions of the $N$ particles on $\Sigma$
is given by a density
$R_\Sigma:\Sigma^N\to \R$ such that
\begin{eqnarray*}\text{Prob(particle $i$ crosses } \Sigma
\text{ in }\delta x_i, \  i=1\dots N) & & \\ =
R_\Sigma(x_1,\dots ,x_N)\delta x_1
\cdots \delta x_N .& & 
\end{eqnarray*}  By $\delta x$ we denote simultaneously an
infinitesimal region on
$\Sigma$ around $x$ and its area (i.e., 3-volume).  Now we
compare
$R_\Sigma$ evaluated at $(x_1,\dots,x_N)\in \Sigma^N$ with
$R_{\Sigma'}$ evaluated at $(x_1',\dots,x_N')\in (\Sigma')^N$,
where $x'\in\Sigma'$ is obtained {}from $x\in \Sigma$ via
displacement {}from $\Sigma$ to $\Sigma'$ in the normal
direction, see Fig.\ \ref{fig:st}.  Let $\delta x'$ be the
area of the image of the region $\delta x$ under this
correspondence. (Since the projection of the Lorentz  metric
on $\Sigma'$ need not agree with the image, under $x\mapsto
x'$, of its projection on $\Sigma$, $\delta x$ and
$\delta x'$ need not agree.)

\begin{figure}
\begin{center}
\epsfxsize=8.5cm
\epsfbox{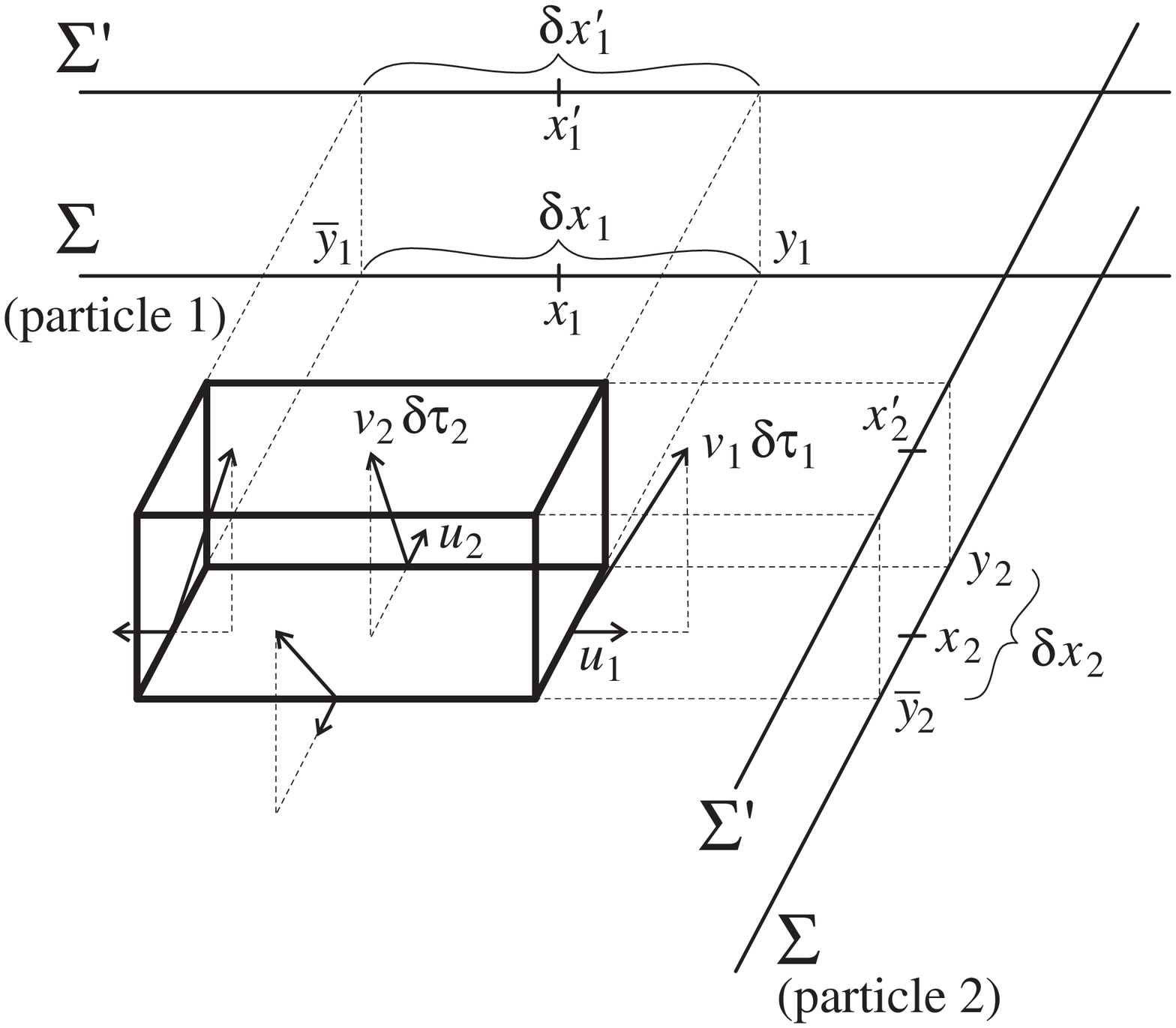}
\end{center}
\caption{Conservation of probability for a system of two
particles in one space dimension:
 configuration-space-time view with, for simplicity, the
hypersurfaces drawn straightened out. (Note that the figure
fails to convey the fact---displayed in Fig.\
\ref{fig:st}---that the areas $\delta x_k$ and
$\delta x_k'$ may differ, and that also $\delta\tau(y_k)$ may
differ {}from 
$\delta\tau(\bar y_k)$, where $y_k$ and $\bar y_k$ are the
boundary points of $\delta x_k$.) The change of the
probability of particle 1 being in $\delta x_{1}$ and particle
$2$ being in
$\delta x_{2}$ {}from hypersurface $\Sigma$ to $\Sigma'$ is
accounted for by the single particle fluxes through the
lateral sides of the configuration space time box between
$\delta x_1\times \delta x_2\subset \Sigma^2$ and the
corresponding set of primed points on $(\Sigma')^2$, i.e., $
R_{\Sigma'}(x_1',x_2')\delta x_{1}' \delta x_{2}' -
R_{\Sigma}(x_1,x_2)
\delta x_{1} \delta x_{2} = $ $ - \left((R_\Sigma v_1) (\bar
y_1,x_2)\cdot (u_1 \delta\tau) (\bar y_1) + (R_\Sigma v_1)
(y_1,x_2)\cdot (u_1 \delta\tau) (y_1)\right)\delta x_{2} $ $ -
\left((R_\Sigma v_2) (x_1,\bar y_2)\cdot (u_2
\delta\tau) (\bar y_2)
 + (R_\Sigma  v_2) (x_1,y_2)\cdot (u_2 \delta\tau) (y_2)
\right)
\delta x_{1} .$  Eq.\ (\ref{eq:qeg}) is the natural extension
of this formula to $N$  particles in Minkowski space.}
\label{fig:cst}
\end{figure}

Recall {}from elementary physics that a continuity equation
such as (\ref{conteqnBD}) is an expression of a local
conservation law that, on the infinitesimal level, can be
stated as follows: The difference between the probability
densities $R_\Sigma$ on $\Sigma^N$ and $R_{\Sigma'}$ on
$(\Sigma')^N$ (with $\Sigma'$ infinitesimally later than
$\Sigma$) is accounted for by the flux through the lateral
sides---to which the hypersurface normals are tangent---of the
configuration-space-time box between $\delta
x_1\times\dots\times \delta x_N\subset \Sigma^N$ and the
corresponding set of (primed) points in $(\Sigma')^N$, see
Fig.\
\ref{fig:cst};
\begin{eqnarray} &  R_{\Sigma'}(x_1',\dots ,x_N') \delta
x_1'\cdots\delta x_{N}'	- R_\Sigma(x_1,\dots ,x_N)
\, \delta x_1\cdots\delta x_{N}	= & \nonumber\\  &  
\displaystyle -
\sum_{k=1}^{N}\delta x_1\ldots\widehat{\delta x_k}\cdots
	\delta x_{N} \int_{\partial (\delta x_k)} (R_\Sigma  v_k)
 (x_1,\dots,x_{k-1},y,x_{k+1},\dots ,x_N)\cdot (u_k
\delta\tau)(y) \, d  S_k ,& 
\label{eq:qeg}
\end{eqnarray} where the $\widehat{\ }$ on $\widehat{\delta
x_k}$ indicates that this term should be omitted {}from the
product. Here $y$ is the integration variable on
$\partial (\delta x_k)$, the (2-dimensional) boundary of
$\delta x_k$ regarded as a region in $\Sigma$, $d S_k $ is the
area element of
$\partial (\delta x_k)$, $u_k$ is the outward unit normal
vector field in
$\Sigma$ to
$\partial (\delta x_k)$, $\delta \tau(y)$ is the Minkowski
distance between
$y\in\Sigma$ and the corresponding $y'\in\Sigma'$ (so that
$y'=y+\delta
\tau(y)n(y)$) and
\begin{equation} v_k= \frac{j_{k}}{j_k
\cdot n_{k}}\label{eq:dvk}
\end{equation} is the covariant velocity of the $k$-th
particle relative to $\Sigma$, see Fig.\ \ref{fig:st}.

Equation (\ref{eq:qeg}) is the continuity equation for the HBD
model in the ``infinitesimally integrated form.'' It is valid
for any hypersurface dynamics defined by (\ref{eq:lawHBD}),
regardless of whether the currents
$j_k$ satisfy (\ref{eq:bas}) and (\ref{eq:divfree}). However,
as we shall now show, if the currents do satisfy
(\ref{eq:bas}) and (\ref{eq:divfree}), then $R_\Sigma= \rho
|_\Sigma = (j_{k}\cdot n_{k})|_\Sigma$ satisfies
(\ref{eq:qeg}). 

Since\footnote{Note that this decomposition is possible
because $\rho$ is defined on $M^N$ (with $M$ Minkowski space),
in contrast with an arbitrary $R=(R_\Sigma)_{\Sigma\in\F}$,
defined only for $N$-tuples belonging to $\Sigma^N$ for some
$\Sigma\in\F$ , for which therefore such a decomposition is
impossible.}
\begin{eqnarray} & & \rho(x_1',\dots ,x_N')\delta
x_1'\cdots\delta x_{N}'- \rho(x_1,\dots ,x_N)\delta
x_1\cdots\delta x_{N}
\nonumber\\ & & = \rho(x_1',\dots ,x_N')\delta x_1'\cdots\delta
x_{N}' \nonumber\\ & & -
\rho(x_1,x_2'\dots ,x_N')\delta x_1\delta x_2'\cdots\delta
x_{N}'\nonumber\\ & & +
\rho(x_1,x_2', \dots ,x_N')\delta x_1\delta x_2'\cdots\delta
x_{N}' \nonumber\\ & & -\rho(x_1,x_2,x_3', \dots ,x_N')\delta
x_1\delta x_2\delta x_3'\cdots\delta x_{N}'
\nonumber\\ & &+\cdots +\rho(x_1,\dots,x_{N-1},x_N')\delta
x_1\cdots\delta x_{N-1}\delta x_{N}' \nonumber\\ & &
-\rho(x_1,\dots ,x_N)\delta x_1\cdots\delta x_{N},
\label{eq:split}
\end{eqnarray} we obtain in this case for the left hand side of
(\ref{eq:qeg}) (to leading order)
\begin{eqnarray} & & \sum_{k=1}^{N}\delta
x_1\ldots\widehat{\delta x_k}\cdots \delta x_{N} \Bigl( j_k
(x_1,\dots , x_k',\dots, x_N)\cdot n(x_k') \delta x_k'
\nonumber\\ & & \qquad  - j_k (x_1,\dots , x_k,\dots, x_N)\cdot
n(x_k)
\delta x_k 
\Bigr),
\label{eq:LHS} \end{eqnarray}  while the integrand on the
right hand side of (\ref{eq:qeg}) becomes
$(j_k\cdot u_k) \delta\tau \, d S_k$. Thus, subtracting the
right hand side of (\ref{eq:qeg}) {}from (\ref{eq:LHS}), we
obtain (to leading order) the  sum over $k$ of the integral of
$j_k$ over the (outward oriented) boundary of the space-time
region above $\delta x_k$ between $\Sigma$ and
$\Sigma'$. But since $j_k$ is divergence-free
(\ref{eq:divfree}), each such term,  and hence  the  sum,
vanishes. Thus (\ref{eq:qeg}) is satisfied, establishing the
equivariance of $\rho$.

We may also write the continuity equation (\ref{eq:qeg}) in a
purely local form: Writing
$$\delta R_\Sigma (x_1,\dots ,x_N) = R_{\Sigma'}'(x_1',\dots
,x_N') - R_\Sigma(x_1,\dots ,x_N),$$ where
\begin{eqnarray} & R_{\Sigma'}'(x_1',\dots ,x_N')\delta
x_1\cdots\delta x_{N} & \nonumber\\ & =R_{\Sigma'}(x_1',\dots
,x_N')\delta x_1'\cdots\delta x_{N}', &  
\label{R'}
\end{eqnarray} and applying Gauss' theorem to the right hand
side of (\ref{eq:qeg})
$$\int_{\partial (\delta x_k)} R_\Sigma v_k
\cdot u_k \delta\tau \, d  S_k = \divv_k^{\Sigma}(R_\Sigma
v_k^{\Sigma}\delta\tau(x_k))\,  \delta x_k,$$ where
$\divv_k^{\Sigma}$ is the divergence with respect to the
$k$-th coordinate $x_k$ on the Riemannian manifold $\Sigma$
and $v_k^{\Sigma}$ is the projection of
$v_k$ on $\Sigma$, yields
\begin{equation}
\delta R_\Sigma  + \sum_{k=1}^{N}\divv_k^{\Sigma}(R_\Sigma
v_k^{\Sigma}\delta\tau_k) =0,
\label{eq:contHBD}
\end{equation} where $\delta\tau_k\equiv\delta\tau(x_k)$.

Using this form we may also check the equivariance of $\rho$.
To do so, we first ``smoothly'' label the hypersurfaces of the
foliation $\F$ by a parameter $s\in \R$, increasing in the
future direction, which may be called a ``time parameter,'' in
terms of which (\ref{eq:contHBD}) becomes a standard
differential equation.  The function $f:M\to \R$ that maps any
point
$x\in M$ to the label $s$ of the hypersurface $\Sigma_s$ to
which $x$ belongs generates the foliation in the manner
described in Section
\ref{sect:surf}. In particular, $\partial f= \|\partial f\|
n$, where $n$ is the future-oriented unit normal vector field
of $\F$.  With $\delta s=
\|\partial f_k\| \delta \tau_k $, where $\partial f_k\equiv
\partial f(x_k)$, we get {}from (\ref{eq:dvk}) that
\begin{equation}v_k\delta \tau_k=  \frac{j_{k}}{j_k
\cdot \partial f_{k}} \, \delta s \equiv \hat v_k \delta s
,\label{eq:vk}
\end{equation} with $\hat v_k= dX_k/ds$ the velocity of the
$k$-th particle in the parametrized formulation of the
dynamics (\ref{equ:dyn}). 

Consider now a coordinate system adapted to our parametrized
foliation
$\Sigma_s$: one coordinate is clearly given by $s$, and on one
foliation hypersurface we introduce an (arbitrary) coordinate
system $p$, which is transported to the other foliation
hypersurfaces by the flow along the normal field, yielding the
system of coordinates $(s,p)$, allowing us to write $x=(s,p)$
for $x\in M$. Then $x_k= (s_k,p_k)\in\Sigma_s$
$\Leftrightarrow$ $s_k=s$, and the relation between $x=(s,p)$
and
$x'=(s',p')$ {}from Fig.\ \ref{fig:st} becomes $p=p'$.  Let
$\delta p$ be the volume element defined by the
$p$-coordinates and let $\delta x=g(p,s)\delta p$. In these
adapted coordinates the continuity equation (\ref{eq:contHBD})
assumes, using (\ref{R'}) and  (\ref{eq:vk}), the more
standard form
\begin{equation}
\frac{1}{g_1\cdots g_N}\frac{\partial (g_1\cdots g_N
R_s)}{\partial s} +
\sum_{k=1}^{N}\divv_k^{\Sigma_s}(R_s
\hat v_k^{\Sigma_s}) =0,
\label{eq:scontHBD}
\end{equation} with $R_s(p_1,\dots ,p_N) =
R_{\Sigma_s}((s,p_1), \dots ,(s,p_N))$ and
$g_k=g(s,p_k),\ k=1,\dots,N $ (and where $\hat v_k^{\Sigma_s}$
is the projection of $\hat v_k$ on $\Sigma_s$).\footnote{This
evolution equation depends upon $g$ only through the
area-expansion factor arising {}from the normal flow between
hypersurfaces, and thus does not really depend upon the choice
of coordinates on the hypersurfaces.}

For $R_s=\rho_s$, (\ref{eq:split}) is what lies behind the
usual (implication of the) chain rule
\begin{eqnarray} & & \frac{1}{g_1\cdots g_N}\frac{\partial
(g_1\cdots g_N
\rho_s)}{\partial s} \nonumber\\ & & =  \sum_{k=1}^N \left.
\frac{1}{g_k}\frac{\partial(g(s_k,p_k)
\rho(s_1,p_1,\dots,s_k,p_k))}{\partial s_k}\right|_{s_k=s} .
\label{eq:drho}\end{eqnarray} Splitting the  4-divergence
into  pieces corresponding to variations orthogonal to and
variations within
$\Sigma_s$, we obtain 
$$ \divv j = \|\partial f\| \left(\frac1{g} \frac{\partial
}{\partial s}
\left(g j^{0} \right) +
\divv^{\Sigma_s} \left(\|\partial f\|^{-1}j^{\Sigma_s} \right)
\right) ,
$$ where $j^0$ is the normal component of $j$, $j^0= j\cdot
n$. Setting 
$j=j_k$ and using $\divv j_k=0$ (\ref{eq:divfree}) we then
find  with (\ref{eq:vk}) that
$$\frac1{g_k}\frac{\partial(g_k \rho)}{\partial s_k}  +
\divv_k^{\Sigma_s} \left(\rho \hat v_k^{\Sigma_s} \right) =0$$
for all
$k$. Therefore, in view of (\ref{eq:drho}), summation over $k$
establishes that
$R_s=\rho_s$ satisfies the HBD continuity equation
(\ref{eq:scontHBD}).

\subsection{Comparison with quantum  mechanics}\label{subs:qm}

The statistical analysis of the hypersurface Bohm-Dirac model
can be based on the assumption that the probability
distribution on $N$-paths is given by the equivariant density
$\rho$ (\ref{eq:covprop}) on some simultaneity surface
$\Sigma$ belonging to the foliation $\F$.  Then, by
equivariance, the statistical predictions of the HBD model
(i.e., the crossing probabilities) agree with the quantum
predictions for positions for any hypersurface in $\F$.  But
what can be said about the statistical predictions concerning
a hypersurface which is not part of a member of
$\F$?

For one particle the situation is very simple: {}From the
geometrical formulation of the HBD model (Section
\ref{sect:surf}) it follows immediately that the HBD model for
one particle is foliation-independent, and in fact is the
usual one-particle Bohm-Dirac theory given by eqs.\
(\ref{BD1particle}) and (\ref{eq:onepDirac}), with  current
$j=\ol\psi\gamma\psi$.  Thus in this case the statistical
predictions of the model agree with the quantum predictions
for position along {\it any}\/ hypersurface.

The situation is analogous for $N$ independent particles: If
the wave function $\psi$ is a product wave function, $ \psi =
\psi_1(\x{1})
\cdots \psi_N(\x{N}) $, then it follows {}from the multi-time
Dirac equation (\ref{Diraceqn}) that $\psi_k$ satisfies the
usual  one-particle Dirac equation.   Furthermore, the path of
the $k$-th particle is tangent to  the one-particle current
$\ol\psi_{k}\gamma\psi_{k}$ and thus   independent of the
paths of the other particles.  Moreover $\rho$ is the product
of the corresponding 1-particle distributions. Therefore, a
product wave function  indeed generates  a
foliation-independent motion,  the  motion of $N$ independent
Bohm-Dirac particles, and we thus have  agreement with all the
quantum position distributions in this case.

In the general case the situation is more subtle: If the
$N$-particle wave function is entangled, it will not in
general be the case that the distribution of crossings of
hypersurfaces not belonging to the foliation agree with the
corresponding quantum position distributions
\cite{Samols,eprli} (which, in fact, may be incompatible with
the crossing statistics for any trajectory model whatsoever). 
However, this disagreement does not entail violations of the
quantum predictions, as has been discussed for the case of the
multi-time translation invariant Bohmian theory in
\cite{eprli}.  In fact, insofar as results of measurement are
concerned, the predictions of our model are the same as those
of orthodox quantum theory, for positions or any other quantum
observables, regardless of whether or not these observables
refer to a common hypersurface belonging to
$\F$.\footnote{This conclusion requires the rather dubious
assumption that the relevant measurements can be understood in
terms of noninteracting Dirac particles. However, in order to
talk coherently about the quantum predictions for a model, it
must be possible to understand measurement processes in terms
of that model. The remarks we are making here would also be
appropriate for the more realistic models for which this would
be true.}  

This is because the outcomes of all quantum measurements can
ultimately be reduced to the orientations of instrument
pointers, counter readings, or the ink distribution of
computer printouts, if necessary brought forward in time to a
common hypersurface in $\F$, or even to a single common
location, for which agreement is assured. Nonetheless, this
situation may seem paradoxical if we forget the non-passive
character of measurement in quantum mechanics. The point is
that for Bohmian quantum theory, measurement can effect even
distant systems, so that the resulting positions---and hence
their subsequently measured values---are different {}from what
they would have been had no measurement occurred.

\section{Perspective}\label{sect:persp}

We have presented a hypersurface Bohm-Dirac model for $N$
entangled but noninteracting Dirac particles.  This model is 
a covariant extension  of the Bohm-Dirac model, which involves
a foliation by equal-time (flat) hypersurfaces, to arbitrarily
shaped (smooth) hypersurfaces.   How natural is this model?

When looking for a relativistic extension of nonrelativistic
Bohmian mechanics one inevitably encounters two central, very
different problems: that such an extension must involve a
mechanism for nonlocal interactions between the particles, and
that quantum equilibrium cannot hold in all Lorentz frames. 
For both of these problems the additional space-time structure
provided by a foliation yields the most obvious solution: The
motion of each particle at a point $x\in M$ depends upon the
paths of the other particles via the points at which they
intersect the leaf of the foliation containing $x$, and we
have an equivariant density on the leaves of the foliation.  

And the simplest way to achieve this, in a covariant manner,
for a Dirac wave function $\psi$, is via the current
(\ref{eq:flhs}): Form the natural tensor $\overline{\psi}
\gamma_1\ldots \gamma_N \psi$, evaluated at $x$ and the other
intersection points, and contract in the slots corresponding
to the other particles with the $N-1$ unit normals to the
hypersurface at the corresponding points, to obtain the
divergence-free 4-vector $j_{k}$, the tangent to the
trajectory at $x$. Thus, \emph{the dynamics of the HBD model
is the simplest Lorentz invariant dynamics compatible with the
structure at hand, namely, the Dirac wave function and the
foliation}. Furthermore, the simultaneous normal component
$\rho=j_k\cdot n_k$ is an equivariant density on the leaves of
the foliation.

It should be stressed, however, that the Lorentz invariance of
the HBD model is---in Bell's sense---``serious'' only if the
foliation is regarded as an additional objective {\it
dynamical}---in contrast to absolute---structure in the theory
(and in the world, if the theory is to describe the world). It
is this structure that is the innovation of what has been
proposed here and in \cite{DGZ90}, not the model per se, which
is indeed a rather straightforward covariant extension of the
BD model.

However, in this paper we shall not try to find a ``serious''
law for the foliation $\F$ or, what amount to the same thing,
its normal vector field
$n$.  As a toy example, however, the foliation law could be
given by an autonomous equation for $n$, such as
$\partial_{\nu} n^{\mu}=0$.  Another class of toy examples
involves a vector field $n$ constructed {}from the wave
function $\psi (\x{1},\dots ,\x{N})$: Consider the space-time
vector fields $v_{kl}^\mu(x) = (\ol\psi \gamma_{k}^\mu\psi)
(\widehat{\x{1}},
\dots ,\widehat{\x{l-1}}, x, \widehat{\x{l+1}}, \dots ,
\widehat{\x{N}})$, where $(\widehat{\x{1}}, \dots ,
\widehat{\x{N}})$ is a point fixed in a
\Li\ way, for example as a maximum of $\ol\psi \psi$.  (Simply
considering
$v_{k}^\mu(x) = (\ol\psi \gamma_{k}^\mu\psi) (x, \dots , x)$
is not a good idea, since this will be zero for antisymmetric
(fermion) wave functions.) Now one may set $n$ equal to the
integrable\footnote{For an arbitrary vector field $v^\mu (x)$,
the Fourier transformed $\hat v^\mu (k)$ may be split into
$\hat v^\mu _\parallel (k) = \hat v^\nu (k) k_\nu k^\mu
/(k_\lambda k^\lambda)$ and $\hat v^\mu _\perp (k) = \hat
v^\mu (k) - \hat v^\mu _\parallel (k)$.  The inverse Fourier
transformed $v^\mu _\parallel (x)$ satisfies the integrability
condition $\partial _\mu v_{\parallel
\nu} - \partial_\nu v_{\parallel \mu} =0$.} part of some
$v_{kl}$.  

A further possibility, which may be more serious, is to have,
in addition to the particle degrees of freedom, an independent
quantum field
$\phi_\mu$ that determines the foliation. Assume that for any
quantum state $\Phi$ of the field, $ (\Phi, \phi_\mu \Phi)$ is
time-like and completely integrable. Then for any state $\Psi$
of the particle-field system, set $ n_\mu = (\Psi, \phi_\mu
\Psi)$. Suppose that the particle and the field degrees of
freedom are both dynamically and statistically independent,
i.e., that there is neither quantum interaction nor
entanglement between these degrees of freedom, so that in
particular the full wave function
$\Psi= \psi \otimes \Phi$.  Then we may define the foliation
by the normal field $n_\mu$.  The $\phi$-field can be regarded
as very roughly analogous to a Higgs field, producing a kind
of spontaneous symmetry breaking, where by choice of $\Phi$ a
particular foliation is determined, and relativistic
invariance thereby broken.

\acknowledgements We thank Folker Schamel for helpful
discussions.   This work was supported in part by the DFG, by
NSF Grant No. DMS-9504556, by Swiss NF grant 20-55648.98, and
by the INFN.

\newpage

\end{document}